\documentstyle{article}

\begin{document}

\title{On the  Design of Cryptographic Primitives}
\date{}
\author{Pino Caballero-Gil$^{(1)}$, Amparo F\'uster-Sabater$^{(2)}$\\
{\small (1) D.E.I.O.C. University of La Laguna. 38271 La Laguna, Tenerife, Spain.}\\
{\small pcaballe@ull.es }\\
{\small (2)Institute of Applied Physics. C.S.I.C. Serrano 144, 28006 Madrid, Spain.}\\
{\small  amparo@iec.csic.es}}
\maketitle
\begin{abstract}
The main objective of this work is twofold. On the one hand, it
gives a brief overview of the area of two-party cryptographic
protocols. On the other hand, it proposes new schemes and
guidelines for improving the practice of robust protocol design.
In order to achieve such a double goal, a tour through the
descriptions of the two main cryptographic primitives is carried
out. Within this survey, some of the most representative
algorithms based on the Theory of Finite Fields are provided and
new general schemes and specific algorithms based on Graph Theory
are proposed.


Classification: 94A60, 11T99,14G50, 11T71
\end{abstract}

\section{Introduction}
\footnotetext{Research supported by the Spanish Ministry of
Education and Science and the European FEDER Fund under Projects
SEG2004-04352-C04-03 and SEG2004-02418.\\
Acta Applicandae Mathematicae. Volume 93, Numbers 1-3, pp. 279-297. Sept 2006. Springer. \\
DOI: 10.1007/s10440-006-9044-3 
}
A two-party cryptographic protocol may be defined as the
specification of a sequence of computations and communications
performed by two entities in order to accomplish some common goal.
For instance, several algorithms may be described in the form of
two-party  protocols, which allow to perform in the
telecommunication world some usual actions such as flipping a
coin,  putting a message in an envelope, signing a contract or
sending a certified mail. This work surveys known protocols based
on finite fields, and proposes new general and specific solutions
based on graphs.

Several approaches to the design of cryptographic protocols have
been carried out from different angles. Some of them have had the
aim of developing a set of standards that can be applied to
cryptographic protocols in general, whereas others have proposed
new specific protocols. The simplest approach to analyze
cryptographic protocols consists in considering them in an
abstract environment where absolute physical and cryptographic
security is assumed. The main disadvantage of this formal approach
is that it does not address potential flaws in actual
implementations of concrete algorithms. On the other hand, the
traditional approach has consisted in guaranteeing the security of
specific protocols based on Finite Mathematics such as the
Quadratic Residuosity Problem and the Discrete Logarithm Problem. In
general such an approach does not allow the composition of
protocols in order to design more complex protocols because it
requires re-modelling the entire system and re-proving its
security. In this paper we propose a mixed approach where security
conditions are guaranteed for certain types of actual protocols.
These algorithms may be used as modules in order to build complex
protocols while maintaining security conditions.

This work is organized as follows. Firstly, basic concepts and
necessary tools are introduced in section 2. Afterwards, specific
notation used throughout the work and general properties of
two-party cryptographic protocols are described in section 3. In
section 4 special attention is paid to the general-purpose
protocol of Oblivious Transfer and its different versions and
applications. Section 5 is devoted to the other primitive of Bit
Commitment and its main application, the so-called Zero-Knowledge
Proof. Finally, several conclusions and possible future works are
mentioned in section 6.

\section{Background}
This work addresses the topic of secure distributed computing
through the proposal of general and specific schemes for some
two-party cryptographic protocols. In such a context, two parties
who are mutually unreliable have to cooperate in order to reach a
common goal in an insecure distributed environment.

The design of cryptographic protocols typically includes two basic
phases corresponding to specification and verification. Up to now,
most works have concentrated on this latter step while a
systematic specification of protocols is almost an undiscovered
area yet. The best known formal methods to analyze cryptographic
protocols that have been published may be classified into three
types. The modal logic based approach is represented by the BAN
logic model for analyzing cryptographic protocols first published
in \cite{BAN90}. On the other hand, one of the earliest works that
used the idea of developing expert systems to generate and
investigate various scenarios in protocol design  was
\cite{CMF87}. A different approach to protocol verification was
based on algebraic systems \cite{DY83}. Regarding research in the
emerging area of formal and systematic specification of
cryptographic protocols, a modular approach was proposed in
\cite{HT94}. Finally, a methodology for both specification and
verification of protocols was presented in \cite{GS95}, and
several basic informal design principles were proposed in
\cite{AN96}.

Note that the design of protocols is not difficult if a Third
Trusted Party (TTP) is available. In such a case, all input
information may be given by both parties to it, and then the TTP
can distribute corresponding outputs to each party. However, the
enormous costs of extra communications, establishment and
maintenance of TTP justify the search for secure non-arbitrated
protocols. In fact, the importance of cryptographic protocol
design lies in the fact that TTP becomes non necessary. Typical
solutions to avoid TTP in cryptographic protocol design include
the use of two powerful tools: computational complexity
assumptions and random choices. In most specific cryptographic
protocol designs the computing power of one or both parties is
supposed bounded. Also usually, some unproven
assumption on the intractability of some finite mathematical problems, and some sort of interaction between both parties are
required. Most two-party cryptographic protocols include the use
of two general techniques, the so-called Cut-and-Choose and
Challenge-Response methods. Cut-and-Choose technique consists in
two stages. First, a party cuts a secret piece of information in several
parts and then the other participant chooses one of them. The goal
of this technique is to achieve a fair partition of the
aforementioned information. The second method, Challenge-Response,
is also formed by two steps. The first step is a challenge from
one of the parties to the other whereas the second step is the
answer to such a challenge.

It is important to remark that most cryptographic protocols are
based on cryptosystems, and therefore, their security depends both
upon the strengths of the underlying cryptosystems, and on the
effectiveness of the protocols in the exploiting these strengths. In particular, most of the protocols analyzed in this work are based on finite mathematical problems that are assumed as difficult in general.
Thus, a poor and careless design may expose protocols to breaches
in security which can be the ideal starting point for various
attacks.

Two-party cryptographic protocols usually consists of series of
message exchanges between both parties over a clearly defined
communication network. Consequently, the possibility always exists
that one or both parties might cheat to gain some advantage, or
that some external agent might interfere with normal
communications. The simplest situation occurs when each party may
function asynchronously from the other party and  make inferences
by combining a priori knowledge with received messages. In a worst
case analysis of a protocol, one must assume that any party may
try to subvert the protocol. As a consequence, when designing a
two-party cryptographic protocol, one of the two following
possible models should be considered. On the one hand, the
so-called semi-honest model is defined when it is assumed that
both parties follow the protocol properly but adversaries may keep
a record of all the information received during the execution and
use it to make a later attack. On the other hand, the so-called
malicious model is considered when it is assumed that different
parties may deviate from the protocol. In order to prove the security in the semi-honest model, the simulation paradigm is
usually applied. According to this paradigm, given the input and
output of any party, it is always possible to simulate through a
probabilistic polynomial time algorithm his or her view of the
protocol without knowing any other input or output. Therefore,
when the simulation paradigm holds, it is obvious that such a
party does not learn anything from the execution of the protocol.
It has been shown \cite{GMW87} that any protocol being secure in
the semi-honest model can be transformed into a protocol that is
secure in the malicious model. This theoretical result has become
an important design principle throughout the field of
cryptographic protocols. It is often easier to start from the
design of a protocol that is secure in a semi-honest model, and
then to transform it into a protocol that is secure in a malicious
model, by forcing each party to prove that he or she behaves as a
semi-honest party.

In conclusion, the security of an interactive protocol should
refer to its ability to withstand attacks by certain types of
enemies. On the other hand, since protocol design is usually based
on the belief that certain computations are difficult, a rigorous
analysis of its security suffers the same limitation.
Consequently, the best that can be hoped for is a demonstration
that either the protocol is secure or that some cryptographic
assumption of the difficulty of a problem is wrong.

\section{Notation and Properties}
Throughout this paper $A$ and $B$ represent the parties Alice and
Bob. The typical notation `$A \rightarrow B$:  $X$' should be
interpreted as `the protocol designer intended $X$ to be
originated by $A$ and received by $B$' because the messages are
assumed not being sent in a benign environment, so there is
nothing in the environment to guarantee that messages are made by
$A$  or received only by $B$.

In order to formalize the notion of cryptographic protocols, $f$
denotes a two-argument finite function from $X_A \times X_B
$  into $ Y_A \times Y_B$, where $X_i$ and $Y_i$ denote private
input and output sets respectively. On the other hand, $s_i$,
$r_i$ and $f_i$ denote three private values corresponding to a
finite value, a random choice and a function respectively. The
sub-indices $i \epsilon \{A,B \}$ indicate the parties $A$ and
$B$.
Thus, a cryptographic protocol may be generally described
through a function $f$ whose public output is defined by  the
expression

$f(x_A,x_B)=f((s_A,r_A),(s_B,r_B))=$

$=(y_A,y_B)=(f_{A}((s_A,r_A),(s_B,r_B)),f_{B}((s_A,r_A),(s_B,r_B)))$.

In this way, at the end of the execution, each party $i \epsilon
\{A,B \}$  receives the output of $f_i$. Note that the previous
definition is independent of the sidedness view of protocols. The
only difference is that in two-sided protocols, both private
outputs are equal.

The security of many cryptographic protocols relies on the
apparent intractability of two mathematical problems known as
Discrete Logarithm Problem (DLP) and Quadratic Residuosity Problem
(QRP) \cite{LN86}. The true computational complexities of these
two problems are not known. That is to say, they are widely
believed to be intractable, although no proof of this is known.
Indeed, both problems are assumed to be as difficult as the
problem of factoring integers. Next the notation corresponding to
these two problems is introduced. Given a prime number $p$, let
$Z_p$ denote the finite field of integers modulo $p$, and let
$Z_p^*$ denote the multiplicative group of integers modulo $p$.
Accordingly, given a composite integer $N$, let $Z_N$ denote the
additive group of integers modulo $N$.

On the one hand, given a prime  $p$, the DLP may be described as a
function from $Z_p^*$ into $Z_{p-1}$. In particular, given a
primitive root $g$ of the finite field $Z_p$, and an integer $y$
between $0$ and $p-1$, the integer $x$ such that $ 0< x < p$ is
referred to as the discrete logarithm of $y$ to the base $g$ if
and only if $g^x \equiv y$ $(mod$ $p)$. The DLP is in $NPI$ class,
which means that no probabilistic polynomial algorithm is known
for solving it. Such a problem has acquired additional importance
in recent years due to its wide applicability in Cryptography
\cite{Odl85}.

On the other hand, given an
odd composite integer $N$, the QRP may be defined as a function from $Z_N$ into $Z_N$.  In particular, given an integer $y$ having Jacobi symbol
$(\frac{y}{N})= 1$, the QRP consists in deciding whether or not
$y$ is a quadratic residue modulo $N$. Note that while the Legendre symbol tells us whether $y$ is a quadratic residue modulo a prime number, the Jacobi symbol cannot be used to decide whether $y$ is a quadratic residue modulo $N$ because if $(\frac{y}{N})= 1$, both cases $y$ being and not being a quadratic residue modulo $N$ are possible.  If $N$ is a product of two
distinct odd primes $p$ and $q$, then $Z_p$ and $Z_q$ are finite fields, and $y$ has no or two square roots. Consequently, in such a case, if the factorization of $N$ is known, the QRP can be solved simply by
computing the Legendre symbol $(\frac{y}{p})$. Conversely, the ability to compute sqare roots modulo $N$ implies the ability to factorize $N$. Otherwise, if the
factorization of $N$ is unknown, then there is no efficient
procedure known for solving the QRP, other than by guessing the
answer.

An advisable methodology for practical
design of cryptographic protocols includes the verification of the following properties.
Firstly, the designer should have a clear idea of what the protocol
should achieve in order to specify the goal, and of what computation and
communication requirements the protocol should satisfy, which
implies the so-called correctness property. However, expressing
the correctness criteria of a protocol is not a trivial task
because most protocols include the use of randomness and
interactions, and are based on some difficult problem or
cryptosystem.
 Also, since the difficulty of the problems and cryptosystems
does not guarantee absolutely the security of the corresponding
protocols, an essential task of their design should be the
anticipation to any possible situation, which corresponds to the
proof of
 fault tolerance (including protection of parties' privacy).
  Finally, cryptographic protocols should be
fair, which means that it should be clearly defined what every
party gets through them.

According to the above comments, it is said that a two-party
cryptographic protocol securely computes a function $f$ in a
semi-honest model, and consequently, that the function $f$ is
securely computable, if the following conditions hold:
\begin{description}
    \item Correctness. Each party may obtain the correct output value of $f$ on those
    input arguments that have been previously distributed between both parties

    \item $i$'s Privacy.  Any value that party $i$ could compute efficiently from certain output of
    $f$, could be computed directly from his or her private input and output.

    \item  Fault-tolerance. The security of the protocol should be stated under any kind of behaviour
from any of both parties or external viewers.

     \item  Fairness.  Both parties should know the full description and possible
 outputs of $f$.
\end{description}

Unfortunately, for most known cryptographic protocols  no results
regarding their correctness, privacy, fault-tolerance and fairness
have been proved. Instead of it, there are security reductions to
prove that the protocols are secure as long as certain
mathematical assumptions are true. Early work on this field
concentrated on privacy as the main security criterion, but later
it was proved to be inadequate since many protocols provide
services that are only indirectly related to privacy.

In the following two sections, the two most important primitives
for the design of cryptographic protocols are analyzed.

\section{Oblivious Transfer}

Oblivious Transfer (OT) is a fundamental two-party protocol that
is used to transfer a secret with uncertainty. It solves the
following situation:   party $A$ knows a secret $s_A$ that wants
to transfer to party $B$ in a probabilistic way such that the following properties are fulfilled:
\begin{description}
  \item Meaningfullness.  $B$ gets the secret $s_A$ with probability $1/2$.
  \item Obliviousness. $B$ knows for sure whether he received the secret $s_A$ but $A$ cannot determine
  whether the transfer was successful any better than random guessing.
\end{description}

Since party $A$ acts only as sender and party $B$ acts only as
receiver, OT is a one-sided protocol. Consequently, it may be
functionally defined as follows:

 $f_B((s_A,r_A),r_B)=s_A$ if $r_A \neq r_B$

It has been formally proved that Secure Two-Party Computation in
general can be reduced to OT in the semi-honest model \cite{IY87},
\cite{BG89} and that OT can be used as a primitive for the design
of any two-party protocol \cite{Kil88}. Furthermore, it has been
established that the existence of one-way trapdoor permutations
guarantees the existence of securely computable OT in the
semi-honest model \cite{Rab81}.

The idea of this definition was first proposed in \cite{Rab81}
where  an algorithm based on the QRP was described. In such a
protocol the secret information to transfer is the factorization
of the product of two large prime numbers. The algorithm may be
described as follows:

\vspace{0.4cm} {\it Rabin OT}
\begin{enumerate}

\item $A \rightarrow B$ the product $N = p q $ of two large prime numbers $p$ and $q$ randomly chosen by herself.

\item $B \rightarrow A$ the integer $x^2$ $(mod$ $N)$, where $x$ is a private random number such that $1 \leq   x \leq N-1$.

\item $A \rightarrow B$ one of the four different square roots \{$x, N-x, y, N-y$\} of $x^2$ $(mod$ $N)$, randomly chosen by herself.

\item If $B$ receives $y$ or $N-y$, then he can compute $p$ and $q$ thanks
to $\gcd((x+y), N)$. Otherwise, he cannot.
\end{enumerate}
\vspace{0.3cm}

After the execution of the previous protocol, $A$ does not know
whether $B$ received the secret or not since her choice was
random. The algorithm uses the fact that the knowledge of two
different square roots modulo $N$ of the same number enables one
to factor $N$. Indeed, from $x^2 \equiv y^2$ $(mod$ $ N)$ we get
$(x+y)(x-y)\equiv 0$ $(mod$ $N)$ and since $x\equiv \pm y$ $(mod$
$N)$, $N$ does not divide $(x+y)$ and does not divide $(x-y)$ yet
it divides $(x+y)(x-y)$, this is only possible if $p$ divides
exactly one of the two terms and $q$ divides the other.
Consequently, through the computation of the greatest common
divisor of $N$ and $(x+y)$, the factorization of $N$ can be easily
computed.

Typical stages and characteristics of OT based on
Challenge-Response and Cut-and-Choose methods are now sketched. In
the following proposed general scheme, $A$'s secret is supposed to
be a solution to a difficult problem, and some complexity
assumption on the computer capacity of both parties is usually
required. The first step implies the definition of a partition of
a difficult instance of the original problem. In the second step
it is required the use of a one-way function $h$ that should have
been previously agreed by both parties. Depending on the
coincidence or difference between both secret random choices
carried out in second and third steps, the transferred solution is
a valid solution to the original difficult problem or not.

\vspace{0.4cm}
{\it General OT}

        \begin{enumerate}

 \item Set-up. $A \rightarrow B$ a partition of an input problem instance $\{P_{0},P_1\}$.
      \item Challenge. $B \rightarrow A$ the output of a one-way function $h$ on a random element
      from one of both sets, $r_B\in P_{j}, j=0\ or\ 1$, $h(r_B)$.
      \item Response. $A \rightarrow B$ the solution to the problem defined by her random choice
      of an element from one of both sets, $r_A\in P_{i}, i \in \{0, 1\}$ and the information sent by
      $B$, $Sol(r_A,h(r_B))$.
      \item Verification.  The secret solution is successfully transferred to $B$ depending on
      both participants' choices.
      \end{enumerate}
\vspace{0.3cm}

According to previous functional definition,
$y_B=Sol(r_A,h(r_B))=s_A$ if $r_A\neq r_B$. Correctness and
privacy properties are satisfied by the General OT due to the
following. $B$ obtains a correct output of function $f$ when both
parties are honest. If $A$ tries to transfer a non-existent secret
solution, a TTP or a Zero-Knowledge Proof (see section 5) might be
used to guarantee correctness. Concerning privacy, after taking
part in the protocol, if $B$ does not receive $A$'s secret
solution, then $B$ cannot obtain it, since his polynomially
bounded computing power does not allow him to solve the problem.
Furthermore, $A$ cannot guess $B$'s secret choice, so she does not
know whether $B$ obtained the secret solution or not.

Next,  a new proposal of OT based on graphs is described. In this
new proposal, which follows the previous general scheme, the
secret to transfer is an isomorphism between two graphs $G_{1}$
and $G_{2}$. The assumption of the following hypothesis is
required: `Computational resources of $A$ allow her to solve the
problem of the isomorphism graphs'.

\vspace{0.4cm}
{\it Graph-Based OT }

\begin{enumerate}
\item Set-up. $A \rightarrow B$ the two graphs $G_{1}$ and $G_{2}$, randomly chosen by herself.
\item Challenge. $B \rightarrow A$ an isomorphic copy $H$ of one of both graphs, $G_{i}$ randomly chosen by himself. \item Response. $A \rightarrow B$ the isomorphism between $H$ and one of the two graphs, $G_{j}$, randomly chosen by herself.

\item Verification.  If the graph chosen by $A$ in the previous step does not coincide with
the one used by $B$ in step 2, $B$ will be able to obtain the
isomorphism between $G_{1}$ and $G_{2}$. Otherwise, $B$ will not
be able to get it.
\end{enumerate}
\vspace{0.3cm}

Note that in Rabin OT, $B$'s choice determines the subsequent
development of the protocol and its security, whilst in our
proposal the security is determined only by $A$'s selection of
graphs.

\subsection{Variants of OT}

The previous description of OT corresponds to its simplest
version. Other two  interesting variations  exist that are known
as one out of two OT (1-2OT) and one chosen out of two OT (1C-2OT)
\cite{EGL82}. The first one is used when $A$ has two secrets and
$B$ wishes to obtain one of them without letting $A$ knows which
one. Thus, 1-2OT may be functionally characterized as follows

$f_B(((s_{A1},s_{A2}),r_A),r_B)=s_{A1}$ if $r_A=r_B$

 (and
otherwise $f_B(((s_{A1},s_{A2}),r_A),r_B)=s_{A2}$).

 The essential difference between a 1-2OT and a 1C-2OT is that in this latter case $B$ is particularly
interested in one of both secrets, so the corresponding functional
definition in this case is

$f_B((s_{A1},s_{A2}),i)=s_{Ai}$.

 In these two variants of OT,
meaningfullness and obliviousness properties should be interpreted
as follows:
\begin{description}
  \item Meaningfullness. $B$ gets exactly one of the two secrets.
  \item Obliviousness. $B$ knows which secret he got but $A$ cannot guess it.
\end{description}

A proof that the three versions of OT are equivalent can be found
in \cite{Cre87}.

Next a known 1-2OT based on the DLP for two secrets $s_0$ and
$s_1$ that are binary strings is described. This algorithm assumes
that both parties $A$ and $B$ know some large prime $p$, a
generator $g$ of $Z_p^*$ and an integer $c$, but nobody knows the
discrete logarithm of $c$.

\vspace{0.4cm}
{\it DLP-Based  1-2OT}
\begin{enumerate}

\item $B \rightarrow A$ two integers $\beta _0 $ and $ \beta _1$ where $\beta _i \equiv g^x$ $(mod $ $ p)$, $\beta _{1-i} \equiv c (g^x)^{-1}$ $(mod $ $ p)$, $i$ is a random bit and $x$ is a random number such that $0 \leq x \leq p-2$.

\item $A \rightarrow B$ the integers  $\alpha _0$ and $\alpha _1$ and the binary strings $r_0$ and $r_1$, where $\alpha _j \equiv g^{y_j}$ $(mod$ $p)$, $\gamma _j \equiv {\beta _j}^{y_j}$ $(mod$ $p)$,  $r_j = s_j XOR \gamma _j$ (without carry), and $y_j$ being integers randomly chosen by herself,  after having checked that $\beta _0 \beta _1  \equiv c$ $(mod $
$p)$.
\item $B$ computes  ${\alpha _i }^x \equiv g^{x y_i} \equiv {\beta _i }^{y_i} \equiv \gamma _i$  $(mod$ $p)$, and  $s_i = \gamma _i  XOR r_i$ (without carry)

\end{enumerate}

\vspace{0.3cm}

Since the discrete logarithm of $c$ is unknown, $B$ cannot know
the discrete logarithm of both $\beta _0$ and $\beta _1$.
Moreover, the information that $B$ sends to $A$ in the first step does no reveal her which of the two discrete logarithms $B$ knows, and consequently which of the two secrets $B$ will receive in the third step.

General OT remains valid for both 1-2OT and 1C-2OT, but with
several modifications. In the case of 1-2OT, the two differences
are in the first and in the fourth step. In most cases, the set-up
step is not necessary. On the other hand, in the verification step
the solution that $B$ receives, $Sol(r_A,h(r_B))$, coincides with
one of both valid solutions depending on $A$'s and/or $B$'s random
choices, $r_A$ and $r_B$.

On the other hand, in 1C-2OT there is a difference in the response
step because there is no $A$'s random choice, and the solution
sent to $B$ is  $Sol(h(r_B))$, where $r_B$ is randomly chosen by
$B$ within the set $P_i$ indicated by his choice.

Next we propose a new solution to this protocol based on the Graph
Theory. In this case, the required hypothesis is that $A$ knows
how to solve a problem $P$ in two graphs $G$ and $H$ and in every
isomorphic copy of them. Also, the graphs $G$ and $H$ are assumed
to have identical polynomially testing properties. The secret to
transfer is now a solution to the problem $P$ in one of two public
graphs $G$ or $H$.

\vspace{0.4cm}
{\it Graph-Based 1-2OT}
\begin{enumerate}
\item Challenge. $B \rightarrow A$ two new graphs:  an isomorphic copy of $G$, $G_{I}$, and an isomorphic copy of $H$, $H_{I}$,
 and a pointer to one of them.
\item Response. $A \rightarrow B$ the solution to the problem $P$ in the graph pointed by $B$ in step 1.
\item Verification.  $B$ transforms the received solution in a solution to the problem $P$ in the original graph $G$
or $H$ by using the isomorphism he knows.
\end{enumerate}
\vspace{0.3cm}

The generalization of 1-2 OT to more than two secrets, known
generally as Secret Sale, is a specially interesting topic  due to
its usefulness in the design of Electronic Elections. The previous
algorithm admits a simple adaptation to a Secret Sale. The only
modification consists in considering $n$ graphs $G_{1}$,
$G_{2}$,\ldots ,$G_{n}$ instead of only two, and an isomorphic
copy $H_{i}$ for each one. Then, the previous outline may be used
for each pair ($G_{i}$, $H_{i}$) so that at the end of the
protocol, $B$  has the solution to the problem on a concrete graph
without allowing $A$ knows exactly in which.

Next, the latter algorithm is used to generate a new OT that
allows to relax the hypothesis of the previous proposal. In order
to do that, we can use the idea of dividing the secret in two
parts so that only if $B$ receives both fractions, he will be able
to obtain the original secret. So, thanks to the composition of
two 1-2 OT, a new OT where the secret is an isomorphism between
two graphs chosen by $A$ may be described. The description of this
new algorithm is as follows.

\vspace{0.4cm}
{\it Graph-Based OT Based on 1-2OT }
\begin{enumerate}
\item Challenge. $B \rightarrow A$ two isomorphic graphs $G_{1}$ and $G_{2}$.
\item Response. $A$ builds an intermediate graph $H$ that is isomorphic to both graphs, and executes two 1-2OT with the isomorphisms between $G_{1}$
and $H$, denoted by $f_{1}$, and between $H$ and $G_{2}$ , denoted
by $f_{2}$.
\item Verification. If both 1-2OT produce the reception of $f_{1}$ and $f_{2}$, $B$ is able to deduce the isomorphism
between $G_{1}$ and $G_{2}$ through the composition of both
received isomorphisms. Otherwise, it is impossible.
\end{enumerate}
\vspace{0.3cm}

\subsection{Applications of OT}

OT has many different and important applications such as Contract Signing, Secret Exchange, Certified Mail, Coin Flipping and Two-Sided Comparison Protocols
\cite{BVV83}. All these applications are analyzed in the following subsections.

\subsubsection{Contract Signing, Secret Exchange and Certified Mail}

The problem of Contract Signing (CS) consists in the simultaneous
exchange between two parties $A$ and $B$ of their respective
digital signatures of a message (contract). The two main
difficulties are to achieve that none of participants can obtain
the signature of the other without having signed the contract and
that none of them can repudiate his or her own signature. This
protocol was first proposed in \cite{Eve82}.

It has been proved \cite{EY80} that no deterministic CS exists
without the participation of a TTP. Thus, since protocols without
TTP are  desirable in distributed environments, CS based on the
use of randomization are specially interesting.

One of the simplest CS is based on the successive application of
Rabin OT. In this case, the contract is considered correctly
signed by both parties if both users at the end know the other
user's secret factors. The same idea can be applied to the
 proposed OT based on graphs, so that the contract will
be signed when both participants have received the other's secret
isomorphism.

A direct relationship exists between CS and two protocols known respectively as
Secret Exchange (SE) and Certified Mail (CM), since all of them
are reducible to each other \cite{DH76}. On the one hand, SE allows that two parties $A$
and $B$ exchange their secrets simultaneously through a
communication network. On the other hand, thanks to a CM party $A$ may
send a message to another party $B$ so that he cannot read it without returning
an acknowledgement of receipt to $A$. In all the three cases of CS, SE
and CM a commitment of exchange of secrets exists that can be
solved by using OT.

Unlike OT, SE is two-sided because both parties act as sender and receiver. It functional definition is as follows:

$f_A(r_A, (s_B,r_B))=s_B$ and  $f_B((s_A,r_A),r_B)=s_A$ if $r_A=r_B$

Next we propose a new Graph-Based SE. Now it is supposed that $A$
knows how to solve the problem of the isomorphism for all the
isomorphic copies of the graph $G_{1A}$, and that $B$ knows how to
solve the isomorphism for all the isomorphic copies of the graph
$G_{1B}$. The secrets to be exchanged are the two isomorphisms
between two graphs $G_{1A}$ and $G_{2A}$, and between the graphs
$G_{1B}$ and $G_{2B}$. Both pairs of graphs are supposed public.

\vspace{0.4cm} {\it Graph-Based  SE}
\begin{enumerate}

\item
\begin{itemize}
\item $A \rightarrow B$ a graph $H_{iB}$ isomorphic copy of one of the two graphs $G_{iB}$, randomly chosen by herself.
\item $B \rightarrow A$ a graph $H_{jA}$ isomorphic copy of one of the two graphs $G_{jA}$, randomly chosen by himself.
\end{itemize}

\item
\begin{itemize}
\item $A \rightarrow B$ the
isomorphism between $H_{jA}$ and one of the two graphs $G^{'}_{jA}$ chosen at random by herself.

\item $B \rightarrow A$ the

isomorphism between $H_{iB}$ and one of the two graphs $
G^{'}_{iB}$ chosen at random by himself.
\end{itemize}
\end{enumerate}
\vspace{0.3cm}

Both steps will be repeated an enough number of times in order to guarantee
that when concluding the execution, the probability that the secrets have not
been mutually exchanged is negligible.

\subsubsection{Coin Flipping}

The main goal of Coin Flipping (CF) is to make jointly a fair
decision between $A$ and $B$, so both users can simulate jointly
the random toss of a coin in a distributed environment. This
protocol has important applications in the generation of secret
shared random sequences in order to use them as session keys in
network communications.

The name of this protocol comes from its first description, given
in \cite{Blu82}. The simplest CF only requires that $A$ and $B$
pick each a random bit $a$ and $b$, and simultaneously exchange
them. In this way, the outcome of the toss may be defined by
$a+b $ $(mod$ $2)$. In order to prevent possible biases of the result,
a Bit Commitment scheme (see next section) might be used. Again,
implementations of CF supported by OT are in general possible. In
order to do it, it is only necessary to specify that the result is
favourable to $B$ when he obtains $A$'s secret, and otherwise it
is favourable to $A$.

A known proposal of CF based on the difficulty of the QRP is shown below.

\vspace{0.4cm}
{\it QRP-Based CF}
\begin{enumerate}

\item $A \rightarrow B$ two integers $N$ and $z$, such that $N$ is a Blum integer (product of two prime numbers $p$ and $q$
that are congruent with 3 modulo 4), and $z\equiv y^2$ $(mod$ $ N)$
with $y \equiv x^2$ $(mod$ $N)$ and $x \in Z_N^*$.

\item $B \rightarrow A$ his bet on that $y$ is even or odd.

\item $A \rightarrow B$ the integers $x$ and $y$, and a proof that $N$ is a
Blum integer.

\item $B$ checks that $y \equiv  x^2$ $(mod$ $N)$ and $z \equiv y^2$ $(mod$ $N)$.
\end{enumerate}
\vspace{0.3cm}

Note that the use of a Blum integer is essential in this scheme
since if both numbers $p$ and $q$ are primes $\equiv 3$ $(mod$ $
4)$, then $-1$ is not a square modulo $p$ and modulo $q$, and it
easily follows that the square function becomes a bijective map
where both the domain and range are the subset of squares in
$Z_N^*$. Consequently, this condition assures that $z$ has not two
square roots with a different parity. As before, $B$ could be
persuaded about $A$'s correct selection of N through a
Zero-Knowledge Proof (see next section).

A General CF based on a trapdoor function and a finite set of
integers is next described. In such a scheme both participants
should agree in advance the trapdoor function $h$, which is
defined on a finite set of integers that contains exactly the same
quantity of odd and even numbers.

\vspace{0.4cm}
{\it General CF}
\begin{enumerate}

\item Set-up. $A \rightarrow B$ the output $y = h(x)$ on an
 element $x \in X$, randomly chosen by herself.

\item Challenge. $B \rightarrow A$ his bet on that $x$ is even or odd.

\item Response. $A \rightarrow B$ the original element $x$.

\item Verification. $B$ checks that $h(x) = y$.
\end{enumerate}
\vspace{0.3cm}

Correctness of the previous scheme is based on the appropriate
choice of the trapdoor function $h$. Thus, if $h$ is not an
injective function, $A$ could know two different values $x$ and
$x'$ with a different parity and such that $h(x) = h(x')$, so in
this case $A$ is not committed to any of both values. On the other
hand, if $h$ can be inverted and it is possible to obtain $x$ from
$h(x)$, it would be feasible to deduce the parity of $x$ from
$h(x)$.

\subsubsection{Two-Sided Comparison Protocols}

Now we consider a new application of OT, which is the problem of
evaluating a specific  function by two parties on secret inputs.
In a general description of a Two-Sided Comparison Protocol (TSCP)
two parties $A$ and $B$ each having a secret ($s_A$ and $s_B$)
want that both learn the finite output of a comparison function
$g(s_A,s_B)$ but none of them learns anything about the other
party's secret. The main characteristic of this two-sided protocol
is that it is a symmetric protocol because both parties do the
same actions and obtain the same result. This general protocol has
important applications in Electronic Voting, Mental Poker and Data
Mining. The main problem of its definition is the simultaneity of
both parties' actions. It has been proved \cite{Yao82} that all
functions with finite domain and finite image can be evaluated
through TSCP.

A general functional  definition of TSCP is as follows:

 $f((s_A,r_A),(s_B,r_B))=(g(s_A,s_B),g(s_A,s_B))$.

The following scheme sketches typical stages and
characteristics of a General TSCP for comparing binary strings,
which is based on a 1C-2OT.

\vspace{0.4cm}
{\it General TSCP}
\begin{enumerate}
\item Set-up. Each party chooses at random $2n$ binary strings of length $k$,
 $\{r^A_{iO},r^A_{i1}\},\{r^B_{iO},r^B_{i1}\}, i=1,2,...,n$
\item Transfer.
\begin{itemize}
\item $A \rightarrow B$ one of the two secrets transferred with
$1C-2OT(r^A_{i0},r^A_{i1})$, chosen according to B's secret binary
string.
\item $B \rightarrow A$ one of the two secrets transferred with
$1C-2OT(r^B_{i0},r^B_{i1})$, chosen according to A's secret binary
string.
\end{itemize}
 \item Computation.
 \begin{itemize}
\item $A \rightarrow B$ the bit-wise addition of all the received strings, with the sum
of A's private strings defined by her secret string
$\sum_i{r^A_{is_i}}$ .
\item $B \rightarrow A$ the bit-wise addition of all the received strings, with the sum
of B's private strings defined by his secret string
$\sum_i{r^B_{is_i}}$ .
\end{itemize}
\item Verification. If both additions are different, both parties deduce the difference between both secret strings.
Otherwise, they do not know anything for sure.
      \end{enumerate}
\vspace{0.3cm}

Note that in the verification step both parties could deduce the
equality of both secret strings if both strings coincide. However,
this would be a probabilistic deduction because the probability to
fail is $2^{-k}$. According to this, the given functional
definition of TSCP is

 if $s_A=s_B$, then
$g(s_A,s_B)=0$ since

$\sum_i{1C-2OT(r^A_{i0},r^A_{i1})+ \sum_i{r^B_{is^B_i}}}=
\sum_i{1C-2OT(r^B_{i0},r^B_{i1})}+ \sum_i{r^A_{is^A_i}}$.

Otherwise, $g(s_A,s_B)=1$, which implies
that no party receives any certain information regarding the
comparison between both secret strings.

Correctness and privacy properties are satisfied by  General TSCP
due to the following. Both parties obtain a correct output of the
function $f$ when they are honest because if both secret strings
coincide, both final sums also coincide. If one of both parties
attempts to do a non-valid 1C-2OT, then a TTP or a ZKP might be
used to guarantee correctness. Concerning privacy, after taking
part in the protocol both parties have received only random
strings that do not allow them to deduce the other party secret
string.

Next three different implementations of the general definition of
TSCP are considered, the so-called  Byzantine Agreement, String Verification  and the Millionaires Problem.

In the protocol described in \cite{SLP82}, and known as Byzantine
Agreement (BA), both parties $A$ and $B$ each having a secret bit,
$s_A$ and $s_B$, want to agree on the same bit, which should be
$s_A=s_B$ if this equality holds. According to this definition, if
one party receives a bit different from the one that he or she
owns then he or she learns the other's bit, but if both parties
receive the same bit that they own, then they do not learn
anything about the other's bit.  That is to say, if $s_A\neq s_B$,
they learn it with probability 1/2, but if $s_A=s_B$ they do not
learn anything.

In this case, the functional  definition of BA is as follows:

 $f((s_A,r_A),(s_B,r_B))=(s_A,s_B)$ if $s_A=s_B$ are identical bits

 (otherwise $f((s_A,r_A),(s_B,r_B))=(r,r)$ where  $r=g(r_A,r_B)$ is a random bit).

General TSCP remains valid for BA by limiting the length of binary
strings to $k=1$. In this way, if $s_A\neq s_B$ both sums coincide
with probability 1/2.

String Verification (SV) may be seen as a generalization of a
BA to binary strings. In this protocol proposed
in \cite{NFW96}, both parties $A$ and $B$ each having some secret
$n$-bit string want to verify whether both strings are equal or not, but
nothing more than that. The functional  definition of SV is as follows:

 $f(s_A,s_B)=(0,0)$ if $s_A=s_B$ are identical strings

 (and otherwise $f(s_A,s_B)=(1,1)$).

Again General TSCP remains valid for SV by considering large
 values of length for the binary strings $k$, so that in the
verification step both parties could deduce the equality of both
secret strings if both strings coincide with an almost null
probability to fail, $2^{-k}$.

In the protocol proposed in \cite{Yao82}, known as Millionaires
Problem (MP), two parties $A$ and $B$ are supposed to be two
millionaires  who wish to know who is richer without revealing any
other information about each party's worth. The functional
definition of this protocol is as follows:

 $f(s_A,s_B)=(0,0)$ if $s_A>s_B$ (and otherwise $f(s_A,s_B)=(1,1)$).

This third version of TSCP is different from the previous schemes
in two important questions. First, it is defined on integer values
instead of binary strings. Also, the comparison is not on equality
or difference but on greater or lesser value. Anyway,  General
TSCP may be easily adapted to be used for MP  by considering the
binary representation of both secret integers  $s_A$ and $s_B$,
and by implementing  General TSCP from the most to the least
significant bits (left to right). In this way the algorithm shows
the most significant bit that is different between both secrets,
and determines the desired relationship. However, note that
according to this suggested implementation, a lower bound on the
difference between both secrets is being transferred.

\section{Bit Commitment}

Bit Commitment (BC) is a two-party cryptographic protocol that is

used to simulate the two main characteristics of an envelope:
\begin{description}
    \item Unalterability. $A$ cannot modify its content once she has sent it to $B$.
    \item Unreadability. $B$ can neither obtain the committed value inside the envelope nor
    any information about it until $A$ opens it.
\end{description}
The first condition is equivalent to the aforementioned
correctness property, and is generally known as {\it binding}
property of BC. The second condition corresponds to the mentioned
privacy property, and is called {\it hiding} property in BC.

In  the functional description of BC, the use of a trapdoor function
$h$ whose inversion is only possible for $A$ is required:

$f_B(s_A,r_A)=h(s_A,r_A)$

According to the original definition of BC, the committed secret is a
single bit $b$, so it might be considered as a surjective mapping from a
large domain to $\{0,1\}$. Consequently, a bit is considered committed by a
random element in the preimage of the mapping at an output value.
From this point of view, BC may be considered as a special type of hash
function. According to this, the binding
property of BC implies that the corresponding mapping is a
function. On the other hand, BC meets the hiding property if both
distributions of elements in the preimage of zero and elements in
the preimage of one are indistinguishable to $B$.

BC was first defined in \cite{Blu82}. Since then, many interesting
algorithms based on various typical cryptographic tools such as
hash functions, secret keys ciphers, pseudorandom generators,
discrete logarithms or quadratic residues have been proposed.
Also, BC has proved to be very useful as a building block in the
design of larger cryptographic protocols, so it may be considered
the second main primitive of cryptographic protocol design.

The first BC shown below is based on the QRP.

 \vspace{0.4cm} {\it QRP-Based BC}
\begin{enumerate}

\item $A \rightarrow B$ the product $N$ of two distinct large prime numbers $p$ and $q$, and a non-square $y \in Z_N^*$ with Jacobi symbol $(\frac{y}{N})= 1$.

      \item $A \rightarrow B$ an integer $c \equiv r^2 y^b$  $(mod$ $N)$ where $r \in Z_N^*$ is randomly chosen by herself.
      \item $A \rightarrow B$ the primes $p$ and $q$ and the integer $r$.
      \item Verification. $B$ checks the received information.
      \end{enumerate}
\vspace{0.3cm}

There is an efficient deterministic algorithm that allows to
compute the Jacobi symbol $(\frac{y}{N})$ without knowing $p$ and
$q$. The binding property of the scheme is guaranteed because if
$p$ and $q$ are known, it is easy to check whether $y$ is a
square. Indeed, $y$ is a square if and only if $y$ $(mod$ $p)$ and
$y$ $(mod$ $q)$ are squares, and this is true if and only if the
Legendre symbols $(\frac{y}{p})$ and $(\frac{y}{q})$ are equal to
1. Note that $c$ is a square if and only if $b=0$. The hiding
property is guaranteed by the difficulty of the QRP. $B$ needs $p$
and $q$ in order to check that $y$ is not a square. However, if
$A$ wants not to reveal them, she should prove that $y$ is not a
square by a Zero-Knowledge Proof (see next subsection).

The following algorithm is based on the DLP in finite fields. In
this case, $A$'s secret is an integer $x$.

 \vspace{0.4cm} {\it DLP-Based BC}
\begin{enumerate}

\item $A \rightarrow B$ a large prime $p$ and a generator $g$  of
$Z_p^*$.
      \item $A \rightarrow B$ an integer $y \equiv g^x$  $(mod$ $p)$ with $1<x<p-1$.
      \item $A \rightarrow B$ the integer $x$.
      \item Verification. $B$ checks the received information.
      \end{enumerate}
\vspace{0.3cm}

The hiding property of the scheme, that is to say, the secret $x$,
is protected by the difficulty of the DLP in finite fields. On the
other hand, the binding property is also hold due to the
following. Since $g$ is a generator of $Z_p^*$, it is not possible
to find another integer $x'\neq x$ such that $1<x'<p-1$ and  $y
\equiv g^{x'}$ $(mod$ $p)$. Consequently, it is important that $g$
is really a generator of $Z_p^*$, and party $A$ should prove it to
$B$ through a Zero-Knowledge Proof (see next subsection).

In most known BC, $B$ is supposed polynomially bounded. Usually
$A$ knows a secret solution to a difficult problem that uses to
commit to a secret bit $s_A$. A general scheme for BC based on the
Cut-and-Choose technique is next proposed:

\vspace{0.4cm}
{\it General BC}
\begin{enumerate}

\item Set-up. $A \rightarrow B$ a partition of an input problem instance $\{P_{0},P_1\}$.
      \item Commitment. $A \rightarrow B$ the witness $h(s_A,r_A)$ obtained through a trapdoor
      function $h$ on a random element $r_A \in P_{s_A}$, where $s_A=b \in \{0, 1\}$.
      \item Opening. $A \rightarrow B$ the secret $s_A=b$.
      \item Verification. $B$ checks the received information.
      \end{enumerate}
\vspace{0.3cm}

The binding property is satisfied by General BC because if $A$
modifies the commitment, then the fraud is detected by $B$ in the
verification step. On the other hand, the hiding property is
guaranteed through the one-way transformation used in commitment
step.

As we may deduce from both proposed general schemes, there are
many coincidences between General OT and General BC. However, in
this latter case, $B$'s role is passive because he is limited to
check the received information in the last verification step.
Consequently, BC may be considered a non interactive protocol
since all the communications are one-way from $A$ to $B$.

Again we propose a new algorithm for BC based on graphs.  In this
case, the committed secret is an isomorphism between two graphs
$G$ and $H$.

\vspace{0.4cm}
{\it Graph-Based BC}
\begin{enumerate}

\item Set-up. $A \rightarrow B$ two non isomorphic graphs $G$ and $H$.

\item Commitment. $A \rightarrow B$ an isomorphic copy of

\begin{enumerate}
\item $G$, if $b = 0$

\item $H$, if $b = 1$.
\end{enumerate}

\item Opening. $A \rightarrow B$ the secret isomorphism.

\item Verification. $B$ obtains $b$ and checks the received isomorphism.
\end{enumerate}

This proposal fulfills both binding and hiding properties.

\subsection{Application of BC: Zero-Knowledge Proof}

The most important application of BC is on the design of two-party
cryptographic protocols known as Zero-Knowledge Proofs. A
Zero-Knowledge  Proof (ZKP) is an interactive two-party
cryptographic protocol that allows an infinitely powerful prover
$A$ to convince a probabilistic polynomial time verifier $B$ about
the knowledge of some secret information without revealing
anything about it \cite{GMR85}.  According to the previous
definition, ZKP has two possible results: to accept or to reject
the proof. The secret information could be a proof of a theorem, a
factorization of a large integer, a password or anything
verifiable, that is to say, such that there is an efficient
procedure for checking its validity. ZKP has proven to be very
useful both in Complexity Theory and in Cryptography. In this
latter subject it has played a major role in the design of strong
identification schemes \cite{FFS88}.

The functional definition of ZKP is as follows:

$f_B((s_A,r_A),r_B)=0$ if $B$ accepts the proof,

(and otherwise
$f_B((s_A,r_A),r_B)=1$).

Three characteristic properties of ZKP are completeness (if the
claim is valid, then $A$ convinces $B$ of it with very high
probability), soundness (if the claim is not valid, then $B$ is
convinced of the contrary with very small probability), and
zero-knowledge ($B$ does not receive any other information except
for the certainty that the claim is valid). This latter property may
be checked through the demonstration that the prover $A$ can be
replaced by an efficient (expected polynomial time) simulator,
which generates an interaction indistinguishable from the real
one. This property is usually proved through a constructive
specification of the way the simulator proceeds. The main
difficulty of this proof is to achieve that the simulator convince
the verifier about the knowledge of the secret information without
actually having it. This problem is usually solved thanks to the
rewinding capability of the simulator, which may use several tries
to answer the verifier without letting him know how many tries the
simulator has used.

Two variants of zero-knowledge may be distinguished depending on
the assumed computing power of possible dishonest parties.
Computational zero-knowledge arises when it would take more than
polynomial time for a dishonest verifier to obtain some
information about the secret, whereas perfect zero-knowledge
involves that even an infinitely powerful cheating verifier could
not extract any information. Both previous notions can also be
characterized through the amount of computational resources
necessary to distinguish between the interaction generated by the
simulator and the verifier, and the one associated to the prover
and the verifier. The existence of computational zero-knowledge
has been proven for any $NP$-problem under the assumption that a
one-way function exists \cite{GMW87}, so it is natural that most
known ZKP are computational ZKP. On the other hand, a
demonstration that the existence of perfect zero-knowledge for an
$NP-complete$ problem would cause the Polynomial Time Hierarchy to
collapse has been given \cite{For87}. These two important results
imply that any time a message is sent, it may be accompanied with
a computational ZKP of that the message is correct, which is
applicable in general to protect distributed secure computation
against malicious parties.

In the following ZKP based on the QRP, \cite{FFS88} the existence
of a TTP is assumed. The only purpose of such a TTP is to publish
a modulo $N$ that is  the product of two secret primes $p$ and
$q$. Again, computations are performed in $Z_N$The secret information chosen by $A$ consists of an integer
$s$ such that it is relatively prime with $N$ and such that
$0<s<N$.

\vspace{0.4cm} {\it QRP-Based  ZKP}
\begin{enumerate}

\item $A \rightarrow B$ an integer $v \equiv s^2$ $(mod$ $N)$.
      \item The following steps are independently iterated $m$ times:
      \begin{enumerate}
      \item  $A \rightarrow B$ an integer $a \equiv x^2$ $(mod $ $N)$, where $x$ is any secret integer such that $0<x<N$.
      \item  $B \rightarrow A$  a random bit $r_B$.
      \item $A \rightarrow B$ the integer $y \equiv x s^{r_B}$ $(mod $ $ N)$.
      \item $B$ checks that $y\neq 0$ and $y^2\equiv a v^{r_B}$ $(mod$ $N)$.
 \end{enumerate}
      \end{enumerate}
\vspace{0.3cm}

If $A$ knows $s$, and both $A$ and $B$ follow the protocol properly, then the response $ y \equiv x s^{r_B}$ $(mod $ $ N)$ is a square root of $ a v^{r_B}$, and consequently the verification condition of the last step
holds because $y^2 \equiv x^2 \equiv av^0$ $(mod$ $N)$ and $y^2 \equiv
x^2s^2\equiv  av^1$ $(mod$ $N)$. Note that $B$ gets no information
about $A$'s secret and in fact, $B$ could play both the roles of
$A$ and $B$. Consequently, the zero-knowledge property is
satisfied.

BC, interactive challenge-response, and
cut-and-choose techniques are
 basic ingredients of ZKP. In general, $A$ `cuts' her secret solution in several parts, commits
to them, and afterwards $B$ chooses at random one of those parts
as a challenge. Some of $A$'s possible responses prove $A$'s
knowledge of the secret solution, whereas the others guarantee
against $A$'s possible fraud. Also typically, ZKP consists of
several iterations of an atomic subroutine. By repeating it an
enough number of times, the verifier's confidence in the prover's
honesty increases. Thus, the number m of iterations should be
agreed by $A$ and $B$ according to their different interests. By
using all previously mentioned ideas, the following general scheme
is proposed in order  to describe most known schemes.

\vspace{0.4cm}

{\it General  ZKP}
\begin{enumerate}

\item Set-up. $A \rightarrow B$  a partition of an input problem instance $\{P_{0},P_1\}$.
      \item Iterations. The following steps are independently iterated $m$ times:
      \begin{enumerate}
      \item Commitment. $A \rightarrow B$ a witness associated to a solution of a random instance
      $r_A$, obtained through a one-way function $h$, $h(r_A)$.
      \item Challenge. $B \rightarrow A$ a random bit $r_B$.
      \item Response. $A \rightarrow B$ the solution to the problem $P_j, j \in \{0,1\}$,
      defined from both random choices
      $r_A$ and $r_B$, and $A$'s secret $s_A$, $Sol(r_A,r_B,s_A)$ .
      \item Verification. $B$ checks the received information.
 \end{enumerate}
      \end{enumerate}
\vspace{0.3cm}

In a ZKP defined according to this general scheme, correctness
is guaranteed through completeness and soundness properties.
Completeness guarantees correct execution of the protocol when
parties act correctly, whereas soundness protects $B$ against a
dishonest party $A$ who does not know the secret. On the other hand,
privacy is reached through zero-knowledge, because this property
assures  that $B$ does not receive any information on the secret
thanks to his participation in the protocol.

The new ZKP described below is based on the primitive of BC. In
the first step of each iteration $A$ commits to her secret
information, which is  a solution to a difficult problem in a
graph $G$. In the verification phase $B$ checks that the
commitment has not been broken. The resulting proposal is a
general method that can be adapted to be used with different graph
problems \cite{CH01}.

\vspace{0.4cm}
{\it Graph-Based ZKP}
\begin{enumerate}
\item Set-up. $A \rightarrow B$ a graph $G$, which is used as her public identification
\item Iterations. The following steps are independently iterated $m$ times:
\begin{enumerate}
      \item Commitment. $A \rightarrow B$ an isomorphic copy  $G'$ of the original graph $G$ where she knows a solution to  a difficult problem.
      \item Challenge. $B \rightarrow A$ a random bit $r_B$.
      \item Response. $A \rightarrow B$ one of the two messages:

         i) the isomorphism between both graphs $G$ and $G'$, if $ r_B = 0$.

            ii) the solution in the isomorphic graph $G'$, if $ r_B = 1$.

      \item Verification. $B$ checks:

         i) the received isomorphism, if $ r_B = 0$.

         ii) that the received information verifies the properties of a
solution in the isomorphic graph $G'$, if $ r_B = 1$.
 \end{enumerate}
\end{enumerate}
\vspace{0.3cm}

The security of this algorithm is based on the difficulty of the
used graph problem and on the choice of both the graph $G$ and the
secret solution. It is also only applicable when the computational
capacity of the verifier is polynomial.

\section{Conclusions}

One of the main objectives of this work has been to provide a
short survey of the two most important primitives in two-party
cryptographic protocols design. Such a review has shown that finite fields play a crucial role in the design of well-known cryptographic protocols.  On the other hand,
formal characterizations of definitions, general schemes for
such primitives, and descriptions of new algorithms based on Discrete Mathematics have also been given within this paper.

This work has emphasized several aspects
regarding typical cryptographic protocol design such as the
existing  relationship among different primitives, the important function played by certain cryptographic primitives as building blocks
of more complex protocols, the presence of common schemes in
various algorithms, and the use of typical ingredients such as
interaction, randomness, and complexity assumptions in the
definition of most algorithms.

\end{document}